\documentclass[runningheads]{llncs}
\usepackage{graphicx}
\begin{document}
\title{Tactile Perception of Objects by the User's Palm for the Development of Multi-contact Wearable Tactile Displays}
\titlerunning{Tactile Perception of Objects by the User's Palm}
%
\author{Miguel Altamirano Cabrera\inst{1}\orcidID{0000-0002-5974-9257} \and \\
Juan Heredia\inst{1}\orcidID{0000-0001-6024-0485} \and \\
Dzmitry Tsetserukou\inst{1}\orcidID{0000-0001-8055-5345}}
\authorrunning{M. Altamirano Cabrera et al.}
%
\institute{Skolkovo Institute of Science and Technology (Skoltech), Bolshoy Boulevard 30, bld. 1, Moscow, Russia 121205
\email{\{miguel.altamirano,juan.heredia, d.tsetserukou\}@skoltech.ru}}
\maketitle              
\begin{abstract}

The user’s palm plays an important role in object detection and manipulation. The design of a robust multi-contact tactile display must consider the sensation and perception of of the stimulated area aiming to deliver the right stimuli at the correct location. To the best of our knowledge, there is no study to obtain the human palm data for this purpose. The objective of this work is to introduce the method to investigate the user’s palm sensations during the interaction with objects. An array of fifteen Force Sensitive Resistors (FSRs) was located at the user’s palm to get the area of interaction, and the normal force delivered to four different convex surfaces. Experimental results showed the active areas at the palm during the interaction with each of the surfaces at different forces. The obtained results can be applied in the development of multi-contact wearable tactile and haptic displays for the palm, and in training a machine-learning algorithm to predict stimuli aiming to achieve a highly immersive experience in Virtual Reality. 

\end{abstract}
\section{Introduction}

The Virtual Reality (VR) experiences are used by an increasing number of people, through the introduction of devices that are more accessible to the market. Many VR applications have been launched and are becoming parts of our daily life, such as simulators and games. To deliver a highly immersive VR experience a significant number of senses have to be stimulated simultaneously according to the activity that the users perform in the VR environment. 

The tactile information from haptic interfaces improves the user's perception of virtual objects. Some devices that increase the immersion experience in VR were introduced, providing haptic feedback at the fingertips \cite{Chinello2015,Kuchenbecker2008,gabardi2016new,prattichizzoTo,MinamizawaePrattichizzo}.

\begin{figure}[h!]
  \centering
  \includegraphics[width=1 \linewidth]{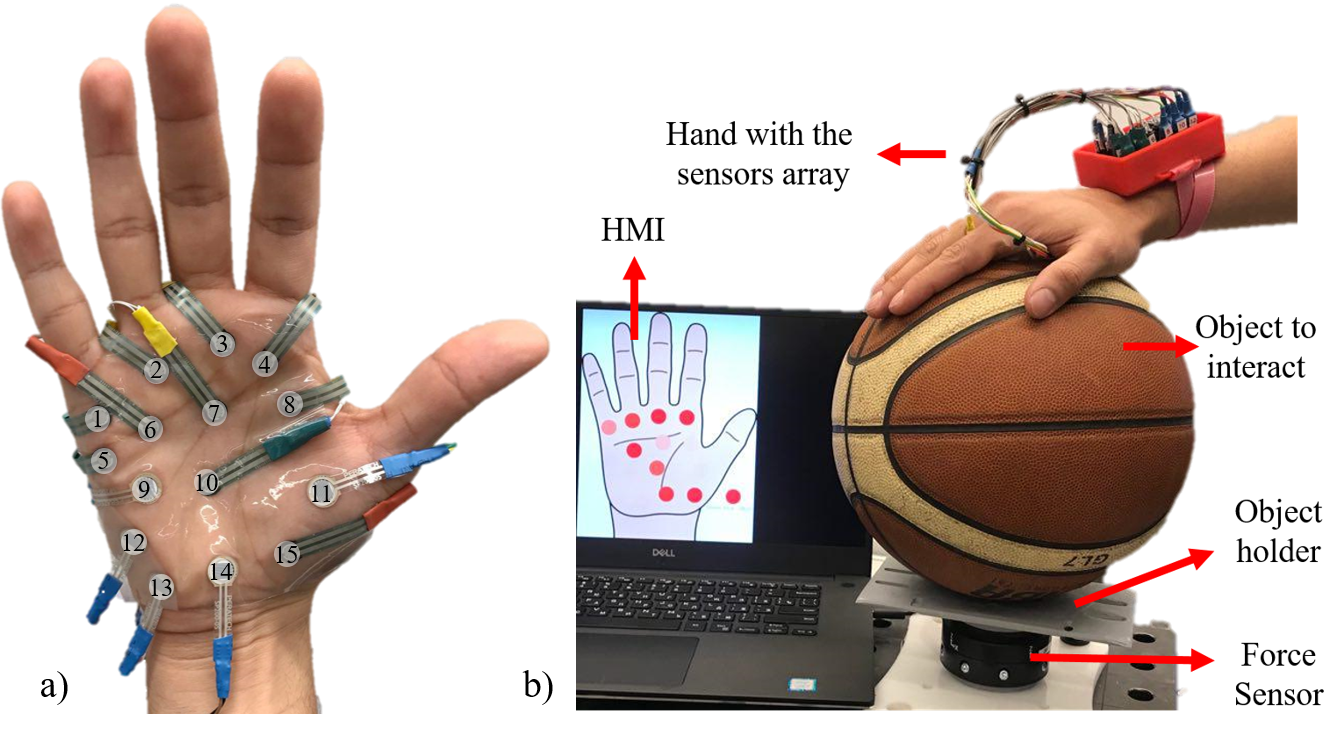}
  \qquad
  \caption{a) Sensor array at the user's palm to record the tactile perception of objects. Fifteen Force Sensitive Resistors (FSRs) were located according to the physiology of the hand, using the points over the bones joints, and the location of the pollicis and minimi muscles. b) Experimental Setup. Each of the four surfaces was located on the holder at the top of the force sensor FT300. The data from the fifteen FSRs and the force sensor are visualized in real-time in a GUI and recorded for the future analysis.}
  \label{fig:handsensors}
\end{figure}

Many operations with the hands involve more that one contact point between the user's fingers, palm, and the object, e.g., grasping, detection, and manipulation of objects. To improve the immersion experience and to keep the natural interaction, the use of multi-contact interactive points has to be implemented \cite{Frisoli2005,Pacchierotti2012}. Choi et al. \cite{Choi2017,Choi2016} introduced devices that deliver the sensation of weight and grasping of objects in VR successfully using multi-contact stimulation. Nevertheless, the proposed haptic display provides stimuli at the fingers and not on the palm.%

The palm of the users plays an essential role in the manipulation and detection of objects. The force provided by the objects to the user's palms determines the contact, weight, shape, and orientation of the object. At the same time, the displacement of the objects on the palm can be perceived by the slippage produced by the forces in different directions. 

A significant amount of Rapidly Adapting (RA) tactile receptors are present on the glabrous skin of the hand, a total of $17,023$ $units$ on average \cite{Johansson1979}. The density of the receptors located in the fingertips is $141\ units/cm^2 $, which is bigger than the density in the palm $25\ units/cm^2$. However, the overall receptor number on the palm is compensated by its largest area, having the $30\%$ of all the RA receptors located at the hand glabrous skin. To arrive at this quantity, we should cover the surface of the five fingertips. For this reason, it is essential to take advantage of the palm and to develop devices to stimulate the most significant area with multi-contact points and multi-modal stimuli. Son et al. \cite{Son2018} introduced a haptic device that provides haptic feedback to the thumb, the middle finger, the index finger, and on the palm. This multi-contact device provides kinesthetic (to the fingers) and tactile stimuli (at the palm) to improve the haptic perception of a large virtual object. 

The real object perception using haptic devices depends on the mechanical configuration of the devices, the contact point location on the user's hands, and the correctness of the information delivered. To deliver correct information by the system introduced by Pacchierotti et al. \cite{PacchierottieKuchenbecker}, their haptic display was configured using a BioTac device, after that, the BioTac device was located in the slave side to perceive the surfaces.  

There are some studies on affective haptics engaging palm area. In \cite{Tsetserukou2010}, the pressure distribution on the human back generated by the palms of the partner during hugging was analyzed. Toet et al. \cite{Toet2013} studied the palm area that has to be stimulated during the physical experience of holding hands. They presented the areas of the hands that are stimulated during the hand holding in diverse situations. However, the patterns extraction method is not defined, and the results are only for parent-child hand-holding conditions.

Son et al. \cite{Son20182} presented a set of patterns that represent the interaction of the human hand with five objects. Nevertheless, the investigated contact point distribution is limited to the points available in their device, and the different sizes of the human hands and the force in the interaction are not considered.  

In the present work, we study the tactile engagement of the user's palms when they are interacting with large surfaces. The objective of this work is to introduce a method to  investigate the sensation on the user's palm during the interaction with objects, finding the area of interaction, and the normal force delivered. This information is used to the reproduction of tactile interaction, and the development of multi-contact wearable tactile displays. The relation between the applied normal force and the location of the contact area should be found, considering the deformation of the palm interacting with different surfaces.

\section{Experimental Setup}
The human palm is a significant area of the hand that plays an essential role in object detection and manipulation. As was mentioned, 30\% of all the RA receptors of the hand are located in the palm \cite{Johansson1979}.

The hands of the users have a different shape, size, and a different elasticity according to their physical activity. Also, during the interaction with the objects or surfaces, the hand experiences deformation according to the applied force. 

A sensor array was developed to investigate the palm area that is engaged during the interaction with the different surfaces. Fifteen FSRs were located according to the physiology of the hand, using the points over the bones joints, and the location of the pollicis and minimi muscles. In Fig. \ref{fig:handsensors} the distribution of the fifteen FSRs is shown.

The fifteen FSRs are held by a transparent adhesive contact paper that is attached to the user's skin. The transparent adhesive paper is flexible enough to allow users to open and close their hands freely. The different shape of the user's hands is considered every time the array is used by a different user. The method to attach the FSRs on the user's palms is the following: a square of transparent adhesive paper is attached in the skin of the users, and the points, where the FSR must be placed, are indicated with a permanent marker. After that, the transparent adhesive paper is taken off from the hand to locate the fifteen FSRs at the right points.

The FSRs are connected to ESP8266 microcontroller through a multiplexer. The microcontroller receives the data from the FSRs and sends it to the computer by a serial communication.

The hand deformation caused by the applied force to the objects is considered in this study. The force applied to the objects is detected by a Robotiq 6 DOF force/torque sensor FT300. This sensor was chosen because of its frequency of $100\ Hz$ for data output and the low noise signal of $0.1\ N$ in $F_z$, which allowed getting enough data for the purposes of this study. The sensor was fixed to a massive and stiff table (Siegmund Professional S4 welding table) using an acrylic base. An object holder was designed to locate the objects at the top of the force sensor FT300.


\section{Experimental Procedure}

The system described in the last section was used in the experimental evaluation to get data from different users and to analyze the relationship between the contact areas at the palm, and the normal force applied. Four surfaces with different diameters were used, three of them are balls, and one is a flat surface. The different diameters are used to focus on the position of the hand: if the diameter of the surface increase, the palm is more open. The diameters of the balls are 65 mm, 130 mm, and 240 mm.  

\textbf{Participants}: Teen participants volunteering completed the tests, four women and six men, aged from 21 to 30 years. None of them reported any deﬁciencies in sensorimotor function, and all of them were right-handed. The participants signed informed consent forms.

\textbf{Experimental Setup}: Each of the four surfaces was located on the holder at the top of the force sensor FT300. The data from the fifteen FSRs and the force sensor are visualized in real-time in a graphical user interface (GUI) and saved for the future analysis.

\textbf{Method}: We have measured the size of the participant's hands to create a sensor array according to custom hand size. The participants were asked to wear the sensors array on the palm and to interact with the objects. Subjects were asked to press objects in the normal direction of the sensor five times, increasing the force gradually until the stronger force they can provide. It was asked to press each object five times. After that, the object was changed, and the force sensor was calibrated.

\section{Experimental Results}

\begin{figure}[h!]
  \centering
  \includegraphics[width=0.7 \linewidth]{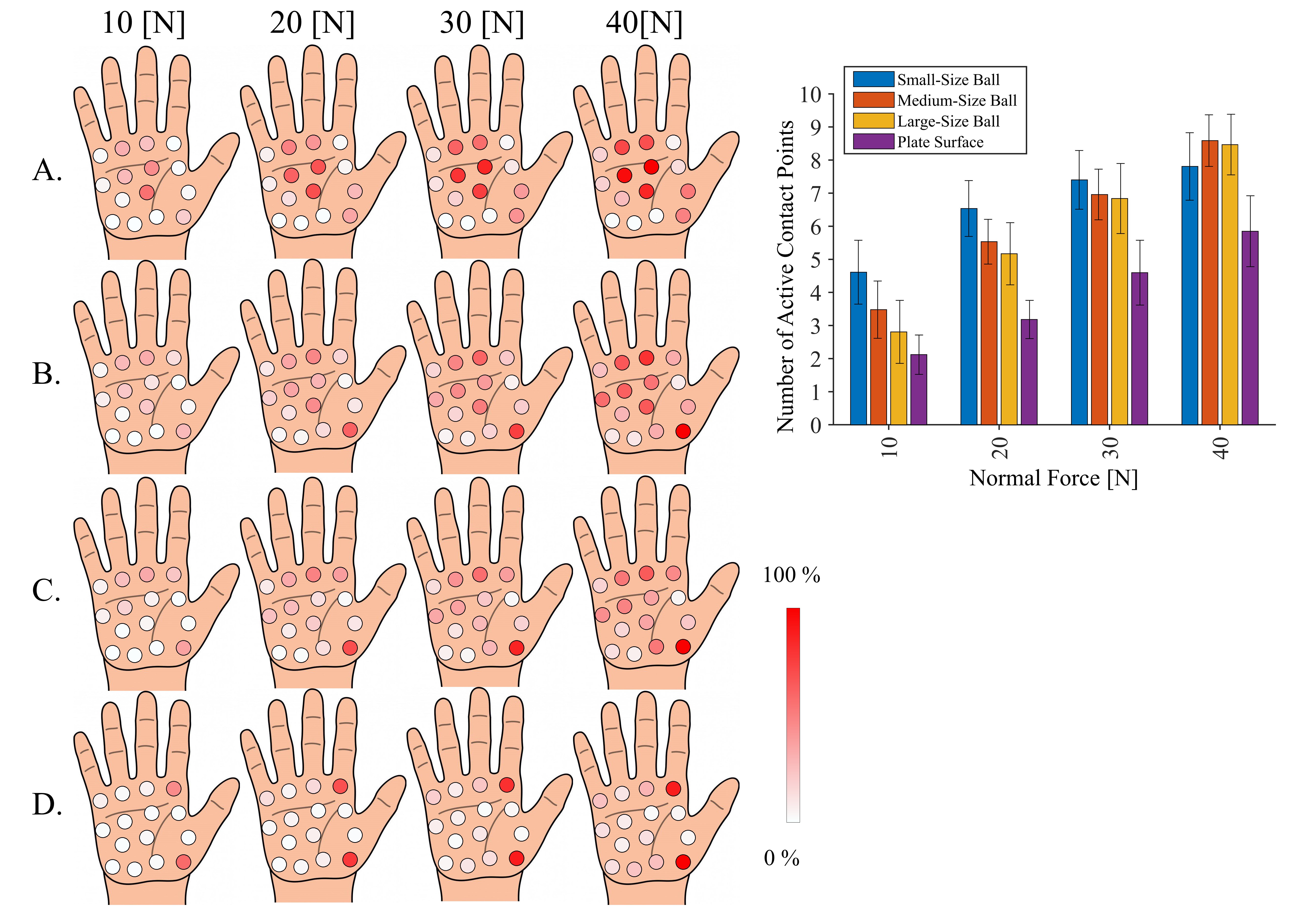}
  \qquad
  \caption{Number and location of contact point engaged during the palm-object interaction. The data from the FSRs were analyzed according to the normal force in each surface. The average values of each FSR were calculated. The active points are represented by a scale from white to red. The maximum recorded value in each surface was used as normalization factor, thus, the maxim recorded value represents $100\%$. In the rows the four surfaces are presented where A is a small-size ball, B is a medium-size ball, C is a large-size ball, and D is a flat surface.}
  \label{fig:handsactive}
\end{figure}

The data obtained in the experimental evaluation were stored for different object interaction (small-size ball, medium-size ball, large-size ball, and flat surface). The force from each of the fifteen FSRs and from the force sensor FT300 was recorded in a rate of 15Hz. To show the results, the data from the FSRs were analyzed when the normal force was 10 N, 20 N, 30 N, and 40 N. In every force and surface, the average values of each FSR was calculated. The active sensors according to the normal force for each surface is presented in Fig. \ref{fig:handsactive}. 

\begin{figure}[h!]
  \centering
  \includegraphics[width=0.60 \linewidth]{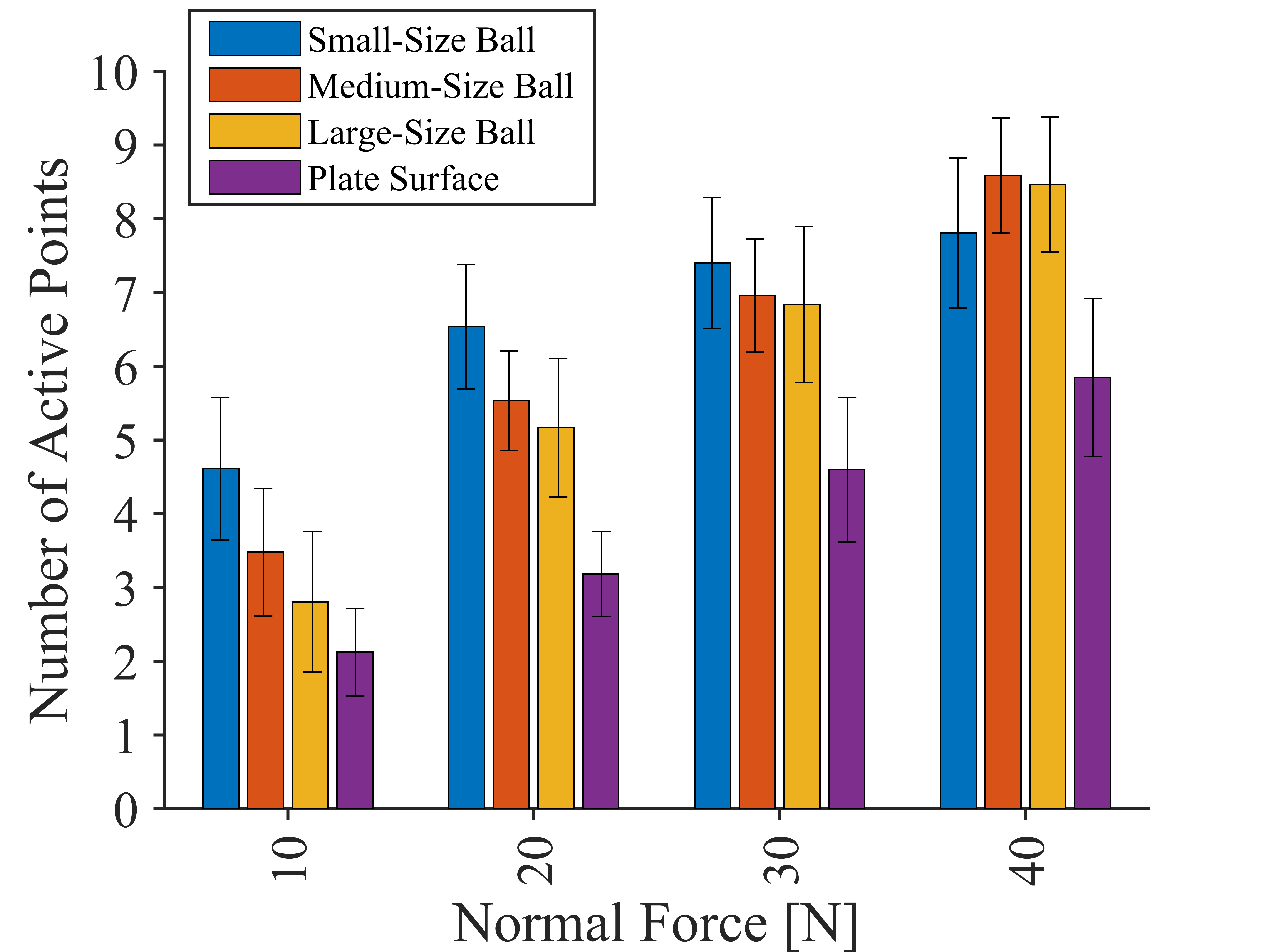}
  
  \caption{Number of contact points in relation with the normal apply force from 10 to 40 N.}
  \label{fig:handsactive6}
\end{figure}

The surface dimension is playing an important role in the activation of FSRs. It can be observed that the surfaces that join up better to the position of the hand activate more points. The sheathe of the palm is related to the deformation to which it is being exposed and to the object surface. When a small-size ball is pressed at 10 N, the number of contact points are almost twice the contact points of other surfaces, but it increased only 60 \% at 40 N. The large-size ball starts at two active contact points at 10 N, and increases to the 150 \% at 40 N. Instead, the flat surface does not activate the points at the center of the palm.

The small-size ball is the most ergonomic adaptable to the hand object. Fig. \ref{fig:handsactive} shows that from small forces, the ball touches many hand points. Contrarily, flat surfaces are not adaptable to hand physiognomy. The contact points are less than on other types of surfaces. However, point 15 and 4, located in the boss down the thumb and the index respectively, are active most of the time. From the results, we can conclude that on flat surfaces, the boss zones contact directly to the object.

\section{Conclusions and Future Work}

The sensations of the users were analyzed to design robust multi-contact tactile displays. An array of fifteen Force Sensitive Resistors (FSRs) was develop for the user's palm, and a force sensor FT300 was used to detect the normal force applied by the users to the surfaces. Using the designed FSRs array, the interaction between four convex surfaces with different diameters and the hand was analyzed in a discrete range of normal forces. It was observed that the hand undergoes a deformation by the normal force. The experimental results revealed the active areas at the palm during the interaction with each of the surfaces at different forces. This information leads to the optimal contact point location in the design of a multi-contact wearable tactile and haptic display to achieve a highly immersive experience in Virtual Reality. Furthermore, some point are not interacting with the sensed object surfaces. 
  
 In the future, we are planing to run a new human subject study to validate the results of this work by a psychophysics experiment. Moreover, we will increase the collected data sensing new objects. With the new dataset and our device, we will design an algorithm capable to predict the contact points of an unknown object.

The result of the present work can be applied to telexistence technology. The array of FSR senses objects, and the interactive points are rendered by a haptic display to a second user. The same approach can be used for affective haptics.

\end{document}